\documentclass[prb,twocolumn,superscriptaddress,showpacs,amscd,amsmath,amssymb,verbatim]{revtex4}

\usepackage{graphicx}
\usepackage{textcomp}
\usepackage[OT1,T1]{fontenc}
\usepackage{color}
\definecolor{red}{rgb}{1,0,0}
\definecolor{blue}{rgb}{0,0,1}

\usepackage{epstopdf}

\begin{document}
\newcommand{\STO}{SrTiO$_3$}

\title{Intrinsic paramagnetism and aggregation of manganese dopants in SrTiO$_3$}
\author{A.~Zorko}
\email{andrej.zorko@ijs.si}
\affiliation{Jo\v{z}ef Stefan Institute, Jamova c.~39, SI-1000 Ljubljana, Slovenia}
\affiliation{EN--FIST Centre of Excellence, Dunajska c.~156, SI-1000 Ljubljana,
Slovenia}
\author{M.~Pregelj}
\affiliation{Jo\v{z}ef Stefan Institute, Jamova c.~39, SI-1000 Ljubljana, Slovenia}
\author{H. Luetkens}
\affiliation{Laboratory for Muon Spin Spectroscopy, Paul Scherrer Institute, CH-5232 Villigen, Switzerland}
\author{A.-K. Axelsson}
\affiliation{London South Bank University, 103 Borough Road, London SE1 0AA, London, UK}
\author{M.~Valant}
\affiliation{University of Nova Gorica, Vipavska 13, SI-5000 Nova Gorica, Slovenia}

\date{\today}
\begin{abstract}
Using local-probe magnetic-characterization techniques of muon spin relaxation and electron spin resonance we have investigated the Mn-induced magnetism of the wide-band-gap perovskite SrTiO$_3$. Our results clearly demonstrate that this diluted magnetic oxide remains paramagnetic down to low temperatures for both doping cases, i.e., when Mn substitutes for Sr or Ti. In addition, both experimental techniques have revealed that the distribution of individual Mn$^{2+}$ and Mn$^{4+}$ ions is nonrandom, as these ions partially aggregate into nanosized clusters.
\end{abstract}
\maketitle

\section{Introduction}

Oxide semiconductors with a fraction of the host cations substituted by magnetic ions, i.e., diluted magnetic oxides (DMOs), have been presently intensively investigated for potential spintronic applications.\cite{Matsumoto,Sharma,Ogale,DietlNV,Dietl,Yamada,Chen,YangAPL}
The quest for new multiferroic/magnetoelectric DMO materials that would be of great technological importance is, however, hampered by the general concern whether their magnetism, coupled or uncoupled to a polar state of the material, is undoubtedly intrinsic.\cite{CoeyMRS, Izyumskaya,Garcia,ValantCM} Lately, there has been huge interest in the electronically and chemically doped wide-band-gap perovskite \STO. In this incipient ferroelectric that is highly polarizable, yet it misses the electric freezing even at $T\rightarrow 0$ due to quantum fluctuations,\cite{MullerSTO} a small concentration of dopants can drastically change its electric and magnetic properties.  To name only the most prominent examples, superconductivity,\cite{Schooley, Reyren} a magnetoelectric multiglass state,\cite{Shvartsman,Kleemann}a two-dimensional electron gas,\cite{Ohtomo, Meevasana} and ferromagnetism\cite{Brinkman, Li, Bert} have been reported in the bulk or at the surface of \STO~and at interfaces with other insulating oxides. 

Manganese impurities in \STO, in particular, have been thoroughly investigated, because they can be multivalent and because of their polar and magnetic moments. They are held responsible for
 a relaxor-type polar state\cite{TkachAPL} when Mn$^{2+}$ substitutes for Sr$^{2+}$ (at the dodecahedral A site), while on the other hand, the material remains quantum paraelectric when Mn$^{4+}$ substitutes for Ti$^{4+}$ (at the octahedral B site).\cite{TkachAM1,ChoudhuryPRB} The polar instability observed in the case of the A-site doping is due to off-center location of the much smaller Mn$^{2+}$ ion with respect to the Sr$^{2+}$ ion and was confirmed by experiments\cite{TkachAM2,Laguta,Levin} and theory.\cite{Kondakova} 
The polar freezing has been further argued to induce a spin-glass state for certain compositions, where a magnetoelectric coupling 
should lead to a novel magnetoelectric multiglass.\cite{Shvartsman,Kleemann} The existence of such a state has been recently hotly debated.\cite{Valant, TkachC, ValantR} As an alternative, the frozen magnetism was suggested to be due to segregation of magnetic species into Mn$_3$O$_4$-like clusters and thus to be extrinsic.\cite{ChoudhuryPRB,Valant} Moreover, such magnetism was shown to critically depend on synthesis conditions.\cite{Valant} 
Thus, the intrinsic magnetic ground state of the Mn-doped \STO~remains elusive and calls for an in-depth magnetic study. Furthermore, the role and size of superexchange interactions between dopants\cite{Kuzian} in \STO~ is unclear. 
These magnetic interactions could further be influenced by a seemingly inevitable presence of oxygen-vacancies ($\rm{V_{O}}$) in these materials,\cite{Middey} which might lead to bound magnetic polarons.\cite{Coey} This mechanism could then explain the ferromagnetic ordering of some Mn-doped \STO~materials,\cite{Middey} but needs to be critically assessed.

In order to address 
all the above-mentioned pending issues about the Mn-doped \STO, we have performed an in-depth local-probe investigation. Here, we combine results of muon spin relaxation ($\mu$SR) and electron spin resonance (ESR) investigations. A further insight into the dopant-induced electronic states in \STO~has been provided by diffuse reflectance spectroscopy (DRS) measurement. We show that all the investigated samples, including the A-site doped ones, remain paramagnetic down to low temperatures and that the Mn ions tend to form dopant-reach islands. Similar nonrandom distributions of dopants on a nanoscale -- spinodal decomposition into regions with high/low concentration of dopants and the same crystal structure as the semiconducting host -- were observed before in different diluted magnetic semiconductors.\cite{Dietl, Criado, Kuroda, Bonanni, BonanniPRL} However, in doped \STO~such aggregation has not been observed yet and is fundamentally different from the recently reported\cite{Yang} tendency of the Mn$^{2+}$ ions to segregate in grain boundaries.

\section{Experimental details}

Powder samples were synthesized according to the synthesis protocols thoroughly explained in Ref.~\onlinecite{Valant}.  
Our study benefits from a systematic synthesis approach employing high-energy homogenization of starting reagents and various different sintering conditions (oxygen partial pressures were varied by applying either O$_2$ or N$_2$ atmosphere in post treatment) thus allowing a production of high-quality powder samples of the cubic perovskite \STO, nominally doped on, both, A and B sites in the doping range from 1 to 3\%. For each doping concentration and a chosen doping site, the O$_2$ and N$_2$ post-treated samples were obtained from the same parent compound.  X-ray diffraction patterns showed no indication of secondary phases in any of the investigated samples. The 3\%-doped samples used in our study were the same as in Ref.~\onlinecite{Valant}. In the following, our samples are labeled by $XcY$, where $X=\rm{A, B}$ denotes the two sites, $c$ the doping concentration in percent, and $Y={\rm O_2, N_2}$ the post treatment atmosphere; e.g. A3O$_2$ stands for the 3\% nominally A-site Mn-doped sample that was post-treated in the O$_2$ atmosphere.

Magnetism on a local scale was inspected with $\mu$SR and ESR. The $\mu$SR data were collected on the General Purpose Surface-Muon (GPS) instrument at the Swiss Muon Source (S$\mu$S), Paul Scherrer Institute (PSI), Switzerland. The measurements were performed in the weak transverse field (wTF) of 3~mT, in zero field (ZF) and in various longitudinal fields (LF) applied along the initial muon polarization in the temperature range $1.6-120$~K. The ESR measurements were performed between 5 and 300~K in the $X$ band (at 9.5~GHz) on a home-built spectrometer. The DRS was performed at room temperature on the Lambda 650 UV-vis spectrophotometer (Perkin-Elmer) equipped with a diffusion sphere.

\section{Results}
\subsection{Muon spin relaxation}
To assess the reportedly intrinsic presence of the frozen spin-glass state in the A-site doped \STO~samples\cite{Shvartsman,Kleemann} on a local scale, we first resort to $\mu$SR, a technique well known for its extreme sensitivity to local magnetic fields.\cite{Yaouanc} The muons, that are initially almost 100\% spin-polarized stop in the sample and experience a local magnetic field $B_\mu$, which leads to a time-dependent muon polarization $P(t)$. Its form reflects a magnitude, distribution and fluctuations of $B_\mu$ and depends of the magnetic coupling of the muon with its surroundings. If the field is rapidly fluctuating, $\nu\geq \gamma_\mu B_\mu$, where $\nu$ denotes the field fluctuation frequency and $\gamma_\mu=2\pi\times 135.5$~MHz/T is the muon gyromagnetic ratio, $P(t)$ will decay monotonically with time in ZF. A quasi-static local field, on the other hand, causes a coherent oscillation of the ZF muon polarization that is damped by slow muon dynamics ($\nu\ll \gamma_\mu B_\mu$) and by a field distribution. The resulting depolarization remains nonmonotonic even for randomly oriented frozen dense (e.g, pertinent to nuclear magnetism\cite{Hayano}) or diluted magnetic moments (e.g, arising from a dilute-alloy spin glass\cite{Uemura}).

\begin{figure}[t]
\includegraphics[trim = 0mm 0mm 0mm 5mm, clip, width=0.85\linewidth]{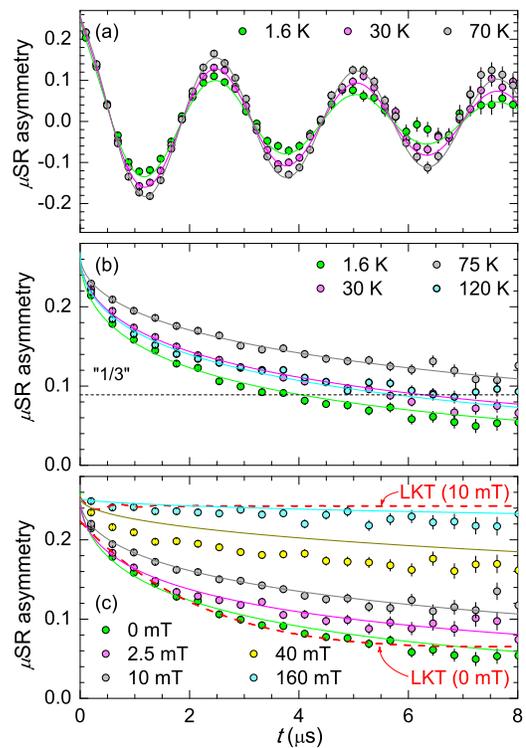}
\caption{The temperature-dependent $\mu$SR asymmetry of the A3O$_2$ sample in (a) wTF and (b) ZF. (c) Field decoupling of the longitudinal $\mu$SR relaxation in various LFs at 1.6~K for the same sample. The solid lines in panels (a), (b) and (c) correspond to fits to the models (\ref{eq4}), (\ref{eq1}) and (\ref{eq3}), respectively. The dashed line in (b) shows the expected position of the "1/3-tail", and in (c) the prediction of the static Lorentzian Kubo-Toyabe model.}
\label{fig1}
\end{figure}

\begin{figure}[t]
\includegraphics[trim = 9mm 5mm 7mm 21mm, clip, width=0.95\linewidth]{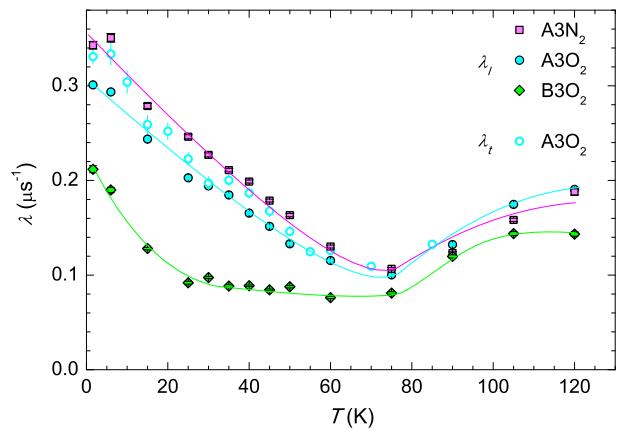}
\caption{The temperature dependence of the longitudinal $\lambda_l$ and transverse $\lambda_t$ muon relaxation rates in Mn-doped \STO. Solid lines are a guide to the eye.}
\label{fig2}
\end{figure}

In Fig.~\ref{fig1}, the $\mu$SR asymmetry $G(t)$ -- spatial asymmetry of muon decay detected by emitted positrons -- that is proportional to $P(t)$ is plotted for the A3O$_2$ sample. The wTF measurements [Fig.~\ref{fig1}(a)] are generally accepted as an effective tool to trace the volume fraction of the frozen spins, as the internal fields are typically well above the applied field and thus cause a loss of the precession amplitude. In diluted spin glasses though, the average internal field may be of the order of only 0.1--1~mT, i.e., much smaller than the applied wTF. However, the spin-glass transition should still manifest in a sharp increase of the muon depolarization rate below the transition temperature, because the on-set of random internal fields broadens the distribution of local fields.\cite{Murnick} Experimentally, we observe a smooth temperature dependence of the transverse relaxation rate $\lambda_t$ in the wTF experiment ({\it vide infra}), with no notable irregularities (Fig.~\ref{fig2}) at the reported spin-glass transition of around 40~K.\cite{Shvartsman,Kleemann,TkachC} The observed damping of the wTF signal can in principle be either due to dephasing in an inhomogeneous static field or due to spin dynamics. These two effects can usually be separated in the ZF experiment. In ZF, the dynamical fields in a diluted magnetic system lead to the monotonic "root-exponential" relaxation\cite{Uemura}
\begin{equation}
G_{\rm ZF}(t,\nu)=G_0{\rm e}^{-\sqrt{4a^2t/\nu}}=G_0{\rm e}^{-\sqrt{\lambda_l t}},
\label{eq1}
\end{equation}
where $G_0$ denotes the initial asymmetry and $a/\gamma_\mu$ the field-distribution width. Static internal fields, on the other hand, in spin glasses lead to a nonmonotonic single-dip relaxation function 
\begin{equation}
G_{\rm ZF}^{\rm LKT}(t)=G_0\left[ \frac{1}{3}+\frac{2}{3}(1-at){\rm e}^{-at} \right],
\label{eq2}
\end{equation}
known as the Lorentzian Kubo-Toyabe (LKT) function.\cite{Uemura}  In the A3O$_2$ sample, both models give a similar quality of fit at 1.6~K [see ZF data in Fig.~\ref{fig1}(c)], with either the dynamical longitudinal relaxation rate $\lambda_l=0.30(1)$~s$^{-1}$ [model (\ref{eq1})] or the static field-distribution width $a/\gamma_\mu=0.32(2)$~mT [model (\ref{eq2})]. We note that in the latter case the dip in $G_{\rm ZF}^{\rm LKT}(t)$ falls outside the experimental time window, therefore, its existence and the leveling of the data at $\frac{G_0}{3}$ at longer times that is characteristic for powder samples\cite{Yaouanc} can not be verified. 

In order to distinguish between the two scenarios, we turn to the LF experiment. In the static case, already the applied field of the order of a few times $a/\gamma_\mu$ noticeably narrows the corresponding local-field distribution, which results in a significantly decreased depolarization. For demonstration, we show in Fig.~\ref{fig1}(c) the prediction of the static LKT model [$a/\gamma_\mu=0.32(2)$~mT] in the applied field\cite{Uemura} of 10~mT, which should almost completely remove the relaxation. This is evidently inconsistent with the experiment and thus unambiguously proves that local magnetic fields at the muon sites are dynamical. Since magnetic fields from nuclei are generally regarded static on the muon time scale, this finding demonstrates that the ground magnetic state of the A3O$_2$ sample is dynamical.
\begin{table}[t]
\caption{XANES-derived occupancy of the Mn$^{2+}$ (Mn$^{4+}$) ions on the A (B) sites in \STO,\cite{Valant} the width of the field distribution at the muon site $a/\gamma_\mu$ and the field fluctuation frequency $\nu$ for the three investigated samples.}
\begin{center}
\begin{ruledtabular}
\begin{tabular}
{c c c c c} & A-site (Mn$^{2+}$) & B-site (Mn$^{4+}$) & $a/\gamma_\mu$ & $\nu$ \\ 
\raisebox{2ex}{sample}&\raisebox{0.7ex}{(\%)}&\raisebox{0.7ex}{(\%)}&\raisebox{0.7ex}{(mT)}&\raisebox{0.7ex}{(MHz)}\\ \hline 
A3N$_2$ & 82 & 18 & 0.67(5) & 4.4(2)\\
A3O$_2$ & 36 & 64 & 0.75(10) & 9.1(5)\\
B3N$_2$ & 13 & 87 \\
B3O$_2$ & 9 & 91 & 0.85(10) & 18.3(4)
\end{tabular}
\end{ruledtabular}
\end{center}
\label{table1}
\end{table} 

In the case of dynamical internal fields with a single fluctuation rate $\nu$, the $\mu$SR asymmetry is given in the applied external field $B_{\rm LF}$ by\cite{Uemura}
\begin{equation}
G_{\rm LF}(t,\nu)=\int_{0}^{\infty}{G_{d}(t,\Delta,\nu)\rho(\Delta,a){\rm d}\Delta}.
\label{eq3}
\end{equation}
Here, $G_{d}(t,\Delta,\nu)={\rm e}^{-\frac{2\Delta^2\nu}{\nu^2+(\gamma_\mu B_{\rm LF})^2}t}$ is the dynamical relaxation function for a field distribution with the width $\Delta/\gamma_\mu$ at a given muon site. In a diluted spin system $\Delta$ is further distributed, which is typically modeled with the Gaussian function $\rho(\Delta,a)=\sqrt{\frac{2}{\pi}}\frac{a}{\Delta^2}{\rm e}^{-a^2/2\Delta^2}$.\cite{Uemura} A simultaneous fit of various LF data sets between 0 and 320~mT [Fig.~\ref{fig1}(c)] to the model~(\ref{eq3}) yields the field-distribution width $a/\gamma_\mu=0.75(10)$~mT and the fluctuation frequency $\nu=9.1(5)$~MHz in the A3O$_{2}$ system at 1.6 K.
We note though that the modeling of the decoupling experiment does not perfectly match the experimental data at intermediate fields [see the 40-mT curve in Fig.~\ref{fig1}(c)], which suggests that there are regions in the sample with different $\nu$. For instance, a region with larger $\nu$ would exhibit faster relaxation than modeled for a single fluctuation rate.

A similar, dynamical behavior was found also in the other two investigated samples: A3N$_{2}$ and B3O$_{2}$. The corresponding field-distribution widths and fluctuation frequencies at 1.6~K obtained from $\mu$SR relaxation in various LFs are summarized in Tab.~\ref{table1}. Both parameters are found to increase with the increasing occupancy of the B site, despite the fact that the average magnetic-moment size decreases -- Mn$^{4+}$ ($S=3/2$) on B sites vs. Mn$^{2+}$ ($S=5/2$) on A sites. 

The temperature dependence of the relaxation rates is shown in Fig.~\ref{fig2} for all three samples. The longitudinal relaxation rates $\lambda_l$ were obtained by fitting the ZF data to the "root-exponential" decay [Eq.~(\ref{eq1})], while the transverse relaxation rates $\lambda_t$ correspond to fits of the wTF data to the model
\begin{equation}
G_{\rm wTF}(t)=G_0{\rm cos}\left( \gamma_\mu B_{\rm wTF} t+\varphi \right){\rm e}^{-\sqrt{\lambda_t t}},
\label{eq4}
\end{equation}
where $\varphi$ denotes a small tilt of the initial muon polarization from the beam direction. First, we stress that  $\lambda_l$ and $\lambda_t$ are nearly equal at each temperature, revealing that the contribution of inhomogeneous static fields to $\lambda_t$ is marginal. This again confirms the dynamical magnetic state in the A3O$_2$ sample at 1.6~K. Second, a minimum of both relaxation rates is observed around 75~K. In the A3N$_2$ sample, the temperature dependence of $\lambda_l$ is very similar, while in the B3O$_2$ sample this minimum is shallower. The increase of the relaxation rates with decreasing temperature below 75~K is typical of slowing-down of spin fluctuations at low temperatures, as $\lambda=\frac{4a^2}{\nu}$.\cite{Uemura} Above 75~K, on the other hand, the increase of the relaxation rates is  rather unusual. 

\subsection{Electron spin resonance}

\begin{table*}[t]
\caption{Parameters of the three ESR components of the Mn-doped \STO~samples, their room-temperature ESR intensities $\chi_{\rm ESR}$ and bulk susceptibilities $\chi_b$ measured by SQUID in the applied field of 10~mT.}
\begin{center}
\begin{ruledtabular}
\begin{tabular}
{c c c c c c c c c c c c c c c} &\multicolumn{4}{c}{Component 1} & \multicolumn{4}{c}{Component 2} & \multicolumn{4}{c}{Component 3}  \\ 
\raisebox{0.5ex}{Sample} &\raisebox{0.5ex}{$I_1$}& \raisebox{0.5ex}{$g_1$}&\raisebox{0.5ex}{$a_1$}&\raisebox{0.5ex}{$\Delta B_1$}&\raisebox{0.5ex}{$I_2$}&\raisebox{0.5ex}{$g_2$}&\raisebox{0.5ex}{$a_2$}&\raisebox{0.5ex}{$\Delta B_2$ }&\raisebox{0.5ex}{$I_3$}&\raisebox{0.5ex}{$g_3$}&\raisebox{0.5ex}{$a_3$}&\raisebox{0.5ex}{$\Delta B_3$} & \raisebox{0.5ex}{$\chi_{\rm ESR}$} & \raisebox{0.5ex}{$\chi_{b}$}\\
 &\raisebox{2ex}{(\%)}& &\raisebox{2ex}{(mT)}&\raisebox{2ex}{(mT)}&\raisebox{2ex}{(\%)}& &\raisebox{2ex}{(mT)}&\raisebox{2ex}{(mT)}&\raisebox{2ex}{(\%)}& &\raisebox{2ex}{(mT)}&\raisebox{2ex}{(mT)}&\raisebox{2ex}{(emu/mol)}&\raisebox{2ex}{(emu/mol)}\\ \hline 
A3N$_2$ & 26(5) & 2.001 & 8.8 & 5.2(3) & 1 & 1.978 & 7.9 & 3.5(5)& 73(5) & 1.998 & / & 32(2)&$3.2(9)\times10^{-4}$&$2.6\times10^{-4}$\\
A3O$_2$ & 27(5) & 2.000 & 8.7 & 6.0(3)  & 1 & 1.982 & 8.0 & 3.5(5)& 72(5) & 1.998 & / & 43(2)&$1.7(5)\times10^{-4}$&$2.3\times10^{-4}$\\
B3N$_2$ & 4(2) & 2.002 & 8.9 & 4.5(3) & 1 & 1.991 & 7.6 & 3.5(3)& 95(2) & 1.998 & / & 35(2)&$1.8(5)\times10^{-4}$&$1.6\times10^{-4}$\\
B3O$_2$ & 3(2) & 2.002 & 8.8 & 5.5(5) & 1 & 1.990 & 7.7 & 3.5(4)& 96(2) & 1.997 & / & 28(2)&$2.5(7)\times10^{-4}$&$2.1\times10^{-4}$
\end{tabular}
\end{ruledtabular}
\end{center}
\label{table2}
\end{table*}  
\begin{figure}[t]
\includegraphics[trim = 0mm 1mm 1mm 1mm, clip, width=1\linewidth]{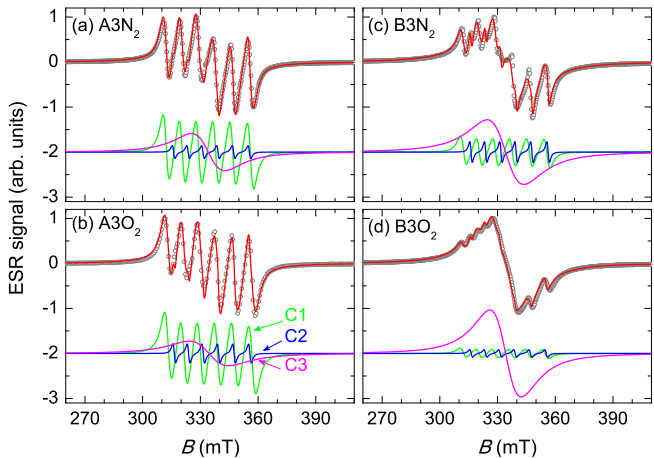}
\caption{The room-temperature ESR spectra of 3-\% Mn-doped \STO~samples (circles) and their fits with a three-component model (see text for details). The three individual components are displaced vertically for clarity.}
\label{fig3}
\end{figure}
In order to obtain further insight into the dynamical magnetic state of the Mn-doped \STO, we next turn to the ESR measurements. ESR has proven highly informative in determining the Mn-ion valence and its local surroundings in \STO. The substitution of Mn$^{4+}$ for Ti$^{4+}$ on the B site was observed by this technique long ago,\cite{Muller} while the successful Mn$^{2+}$ for Sr$^{2+}$ substitution on the A site has been claimed only recently.\cite{Laguta} At high temperatures, both cases are characterized by distinctive hypefine-split ESR sextets, with the $g$ factor $g=1.994$ (2.004) and the hyperfine coupling $a=7.5$~mT (8.8~mT) for Mn$^{4+}$ (Mn$^{2+}$).\cite{Muller, Serway} Such sextets are consistent only with a cubic local crystal-field symmetry and negligible exchange interactions. For the off-center position of the Mn$^{2+}$ ions at the A sites, the absence of the expected single-ion magnetic anisotropy is a consequence of fast ionic motion between energetically equivalent sites.\cite{Laguta} Moreover, in reduced B-site doped \STO~samples, Mn$^{2+}$ was even argued to reside on the B site,\cite{ChoudhuryPRB, Middey} reflecting in the notably different hyperfine coupling $a=8.4$~mT and a significantly reduced $g$ factor.\cite{Middey} Finally, in reduced single crystals, much less intense single-ion-anisotropy split ESR sextets were also observed and assigned to the Mn$^{2+}$-$\rm{V_{O}}$ centers\cite{Serway} and Mn$^{3+}$-$\rm{V_{O}}$ centers\cite{Serway, Azamat} that should locally compensate the excess charge of vacancies.
\begin{figure}[t]
\includegraphics[trim = 1mm 0mm 1mm 2mm, clip, width=1\linewidth]{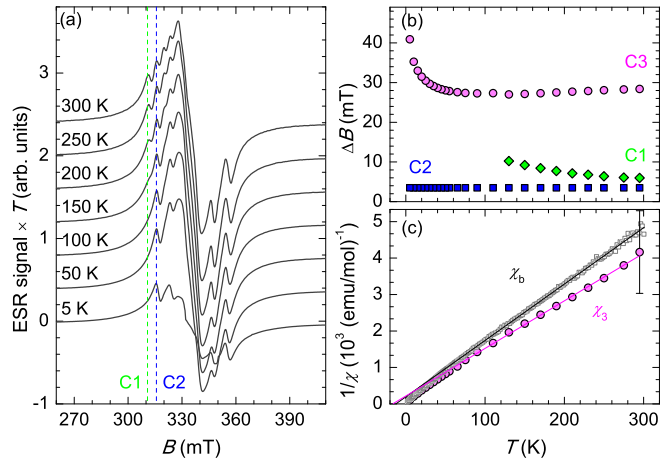}
\caption{Temperature-dependent ESR spectra of the B3O$_2$ sample. The individual spectra are displaced for clarity. The lowest-field peaks of both sextets are marked by dashed lines. (b) The temperature dependence of the ESR line widths of the three components. (c) Comparison of the inverse ESR susceptibility of the broad component $1/\chi_3=1/\left(I_3\chi_{\rm ESR}\right)$ and bulk susceptibility $1/\chi_b$, measured in the field of 10~mT. Solid lines are linear fits of the high-temperature data.}
\label{fig4}
\end{figure}

The room-temperature $X$-band ESR spectra of all four 3\%-doped samples are compared in Fig.~\ref{fig3}. They can be nicely fit to a sum of two Lorentzian-broadened sextets (components C1 and C2) and a much broader C3 component that also exhibits the Lorentzian line shape. 
We note that the Lorentzian shape can either result from motional/exchange narrowing\cite{Bencini} of the absorption line or from the Lorentzian distribution of the local fields in a random dilute system due to $1/r^{3}$-type interactions.\cite{Helm}
The position and the splitting of the observed sextets are characteristic of the Mn$^{2+}$ ions at the A site (C1 sextet) and the Mn$^{4+}$ ions at the B site (C2 sextet);\cite{Muller, Serway} see Tab.~\ref{table2}. We stress that in neither of our samples did we observe any Mn$^{3+}$ resonances, which include an angular independent (-1$\leftrightarrow$1) transition that is in $X$ band expected around 0.6 T.\cite{Azamat} Therefore, we attribute all the observed ESR components to the Mn$^{2+}$ and Mn$^{4+}$ ions. Moreover, we note that our calibration of the ESR intensity with a standard sample revealed a nice correspondence between the ESR susceptibility $\chi_{\rm ESR}$ and the bulk susceptibility $\chi_b$, measured with a SQUID magnetometer (see Tab.~\ref{table2}). This agreement proves that ESR detects all the spins in the Mn-doped \STO~samples and that all three distinct components of the ESR spectrum are intrinsic. Moreover, we find a relatively good correspondence (except for the A3N$_2$ sample) between the relative intensity $I_1$ and the occupancy of the A sites with Mn$^{2+}$ ions (see Tab.~\ref{table1}). Therefore, the broad ESR component is mainly attributed to the Mn$^{4+}$ spins on the B site.
\begin{figure}[b]
\includegraphics[trim = 0mm 0mm 0mm 1mm, clip, width=1\linewidth]{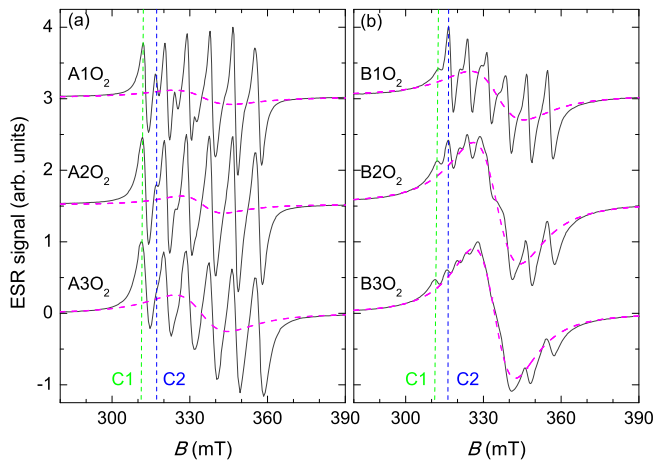}
\caption{Doping-dependent normalized ESR spectra of the \STO~samples (solid lines) at room temperature. The spectra corresponding to different concentrations are displaced for clarity. The thick dashed lines show the broad-line (C3 component) contribution to each spectrum, while the thin vertical lines indicate the lowest-field peaks of both sextets.}
\label{fig5}
\end{figure}

The temperature evolution of the three ESR components is shown in Fig.~\ref{fig4} for the B3O$_2$ sample, in which the broad C3 component is the most intense. The C1 sextet disappears below $\sim$120~K, similarly as observed before.\cite{Laguta} This can be attributed to freezing of the Mn$^{2+}$ ionic motion between the six energetically equivalent off-center positions within the A-site dedecahedron, which results in a much larger magnetic anisotropy and, consequently, in a more complicated anisotropy-split ESR powder spectrum below $\sim$120~K.\cite{Laguta} A gradual slowing down of the Mn$^{2+}$ hopping with decreasing temperature reflects already above 120~K in the increasing line width of the C1 component. The C2 sextet, on the other hand, remains essentially unchanged down to 5~K, with the exception of its intensity that exhibits a Curie-type increase, typical for free magnetic moments. The intensity of the broad component $\chi_3=I_3\chi_{\rm ESR}$ that dominates $\chi_{\rm ESR}$ follows $\chi_b$ [Fig.~\ref{fig4}(c)]. They both yield a small antiferromagnetic Weiss temperature of $\theta=-15(5)$~K. The line-width of the C3 component, on the other hand, is more or less temperature independent down to around 50~K and then increases notably down to 5~K. This line-width increase can be interpreted as a fingerprint of a slowing down of spin fluctuations due to developing spin correlations. Once again, in accordance with the $\mu$SR results, we observe no anomalies in any of the ESR components (and in any of the investigated samples) around 40~K, where the spin-glass state was suggested to set in for the A-site doped samples.\cite{Shvartsman,Kleemann,TkachC} Our measurements thus provide a local-probe verification that such a spin state is not intrinsic and that the Mn-doped \STO~samples rather remain paramagnetic down to the lowest temperatures.

In order to determine the origin of the broad ESR component that is dominant in all samples (see $I_3$ in Tab.~\ref{table2}), we have performed a doping-dependent study for all four different sample types. In Fig.~\ref{fig5}, the ESR spectra in the doping range $c=1 - 3$\% are shown for the families with the least intense (A$c$O$_2$) and the most intense (B$c$O$_2$) broad component. In all cases we observe that the C3 linewidth notably and systematically narrows with the increasing doping concentration, e.g., from 40(2) to 28(2)~mT for the B1O$_2$ and B3O$_2$ samples, respectively. At the same time, its relative intensity increases only slightly (note that the increase of the spectrum height with doping is predominantly due to line narrowing, as the height is proportional to $1/\Delta B^{2}$), mostly at the expense of the C2 sextet, while the relative intensity of the C1 sextet remains unchanged within the experimental error bars.

\section{Discussion}

The intensity of the unstructured broad ESR component C3 dominates the ESR spectra in all investigated samples.  Although such component with a Lorentzian line shape is regularly observed in Mn-doped \STO, especially for B-site doping,\cite{ChoudhuryPRB,Kuzian,Middey,Azzoni} its origin remains unclear. Its alleged absence in some cases, e.g., as in Ref.~\onlinecite{ChoudhurySR}, should be taken with caution, given the fact that due to its broadly distributed spectral weight with respect to the much narrower individual lines within sextets, it can easily be overlooked.
It was ascribed before to the presence of the segregated MnTiO$_3$ phase.\cite{Kuzian}  Since this phase orders magnetically at $T_N=65$~K, causing severe broadening of the ESR line close to $T_N$ and its disappearance below $T_N$,\cite{Stickler} while in our samples the width of the broad ESR line changes gradually and smoothly down to 5~K, this can not be the case here. Next, in reduced SrTi$_{1-x}$Mn$_x$O$_{3-\delta}$ samples, the broad ESR component was correlated with a ferromagnetic response, that was either claimed intrinsic\cite{Middey} or extrinsic.\cite{ChoudhurySR} Again, our samples show no ferromagnetic characteristics.  

In order to clarify its origin and the corresponding relevance for the magnetism of the Mn-doped \STO, we have critically assessed possible ESR broadening mechanisms. As the isotropic exchange interaction $J$ commutes with the $S_z$ operator and therefore yields a delta-function ESR response, only a finite magnetic anisotropy gives a finite ESR line width.\cite{Bencini} First, we consider the case of well-isolated magnetic species ($J\sim 0$). The dipolar interaction between magnetic moments is always present, therefore, we have calculated its contribution to the ESR line width. In the slow modulation limit ($k_B J\ll g\mu_B \Delta B$, where $k_B$, $\mu_B$, and $B_0$ are the Boltzman constant, the Bohr magneton, and the applied magnetic field, respectively) the second moment of the absorption line arising from the secular part of the dipolar interaction,\cite{Bencini}
\begin{equation}
M_2=\left(\frac{\mu_0}{4\pi}\right)^2\frac{3S(S+1)(g\mu_B)^4}{4}\left\langle \sum_k\frac{\left(3\cos^2 \theta_{jk}-1\right)^2}{r_{jk}^6} \right\rangle_j,
\label{eq5}
\end{equation}
yields the ESR line width $\Delta B^d\simeq\sqrt{M_2}/g\mu_B$. Here, the sum runs over all the neighboring sites of a chosen site $j$ at the distance $r_{jk}$ that are occupied by spins $S$, $\theta_{jk}$ is the angle that the ${\bf r}_{jk}$ vector makes with the external magnetic field, and $\left\langle ... \right\rangle_j$ denotes powder averaging. In the case of diluted magnetic lattices, one can sum over all the lattice sites and the dilute-limit line width is then given by $\Delta B^d(c)=c\cdot \Delta B^d$, because $M_2$ is an additive quantity for the $1/r^3$-type interactions.\cite{Uemura} We find the line widths $\Delta B_{\rm A}^d=205$~mT and $\Delta B_{\rm B}^d=134$~mT in the limit of either fully occupied A-sites with Mn$^{2+}$ ions or B-sites with Mn$^{4+}$ ions, respectively. 
In the scenario $(i)$ of diluted dopants, these then yield $\Delta B^d_A(3\%)=6$~mT and $\Delta B^d_B(3\%)=4$~mT, in disagreement with the enhanced width of the C3 component. The scenario $(ii)$ of enhanced dipolar interaction within Mn-rich clusters of \STO~was suggested before to explain the broad ESR component.\cite{ChoudhuryPRB}
 We note though that for dense magnetic clusters the $\Delta B_{\rm A}^d=205$~mT and $\Delta B_{\rm B}^d=134$~mT line widths should be observed. These exceed the experimental ones by an order of magnitude. Moreover, the Gaussian distribution of internal fields would be expected,\cite{Bencini} which is clearly incompatible with the experimental line shape. Alternatively, reduced line widths could occur if the B sites within clusters were only partially occupied; e.g., 22\% magnetic-moment density on B sites would yield the ESR line width of $~$30~mT. However, even partially occupied clusters can not explain the substantial experimentally observed decrease of $\Delta B_3$ with increasing doping concentration. The latter would decrease the distance between magnetic moments and thus increase the average dipolar interactions.  In scenario $(iii)$ the magnetic anisotropy responsible for the width of the C3 ESR component is induced by oxygen vacancies. A vacancy in the B-site oxygen octahedron induces a large uniaxial single-ion anisotropy that was estimated as $D=0.80$~K for the Mn$^{2+}$-$\rm{V_{O}}$ center in reduced \STO~single crystals.\cite{Serway} Since $k_B D/g \mu_B \Delta B_3\sim 20$, isolated Mn$^{2+}$-$\rm{V_{O}}$ centers would yield a complex and much broader ESR spectrum. 

The C3 ESR component thus rather speaks in favor of the scenario ($iv$) where strongly interacting magnetic moments on the B site are responsible for the observed behavior. In this case, large isotropic exchange ($k_B J\gg g\mu_B \Delta B$) leads to exchange narrowing and to a featureless Lorentzian line shape with the line width $\Delta B\simeq M_2/k_B J g\mu_B$.\cite{Bencini} Considering the dipolar interaction within dense clusters (taking into account also the nonsecular terms that renormalize\cite{Bencini} $M_2$ by 10/3) we estimate the required exchange interaction as $J\sim3$~K, while for the single-ion anisotropy scenario the required exchange coupling is of the order $J\sim k_B D^2/g \mu_B \Delta B_3=15$~K. Indeed, the former prediction, yielding the mean-field Weiss temperature $\vert\theta\vert =nS(S+1)J/3\sim 20$~K, where $n =6$ denotes the number of nearest neighbors in a cubic lattice, is in excellent agreement with the experimental value $\theta=-15(5)$~K. 
\begin{figure}[t]
\includegraphics[trim = 0mm 0mm 0mm 5mm, clip, width=0.85\linewidth]{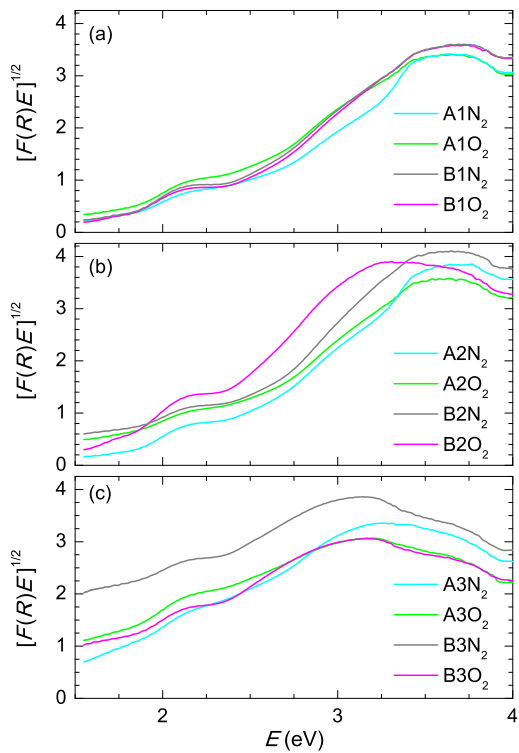}
\caption{The room-temperature diffuse reflectance spectra of (a) 1\%, (b) 2\% and (c) 3\% Mn-doped \STO~samples.}
\label{fig6}
\end{figure}

In principle, the sizable magnetic exchange ($J\sim3$~K) that is needed to account for the broad ESR component could arise from two distinct origins: either from magnetic clusters, where sizable superexchange interaction between nearest-neighbor-occupied B sites was theoretically predicted,\cite{Kuzian} or from a donor-impurity mediated interaction\cite{Coey} in the limit of diluted 
and homogeneously dispersed dopants. In order to further inspect the latter scenario, which stems from oxygen vacancies and has been recently claimed responsible for ferromagnetism of reduced Mn-doped \STO~crystals,\cite{Middey} we have performed DRS measurements, which can provide information on in-gap states for band-gap insulators such as \STO, with the band gap of 3.2 eV.\cite{Benthem}
In Fig.~\ref{fig6} we show the DRS spectra $[F(R)E]^{1/2}$ as a function of the incoming-light energy $E$, which are obtained from the reflectance ($R$) spectra of the samples by the Kubenka-Munk transformation $F(R)=\frac{(1-R)^2}{2R}$.\cite{Kubenka} Similarly as in Ref.~\onlinecite{Middey}, we observed a pronounced dopant-induced in-gap state at around 2~meV that is absent in the undoped \STO~sample.\cite{Middey} This state can be interpreted as a deep vacancy level, which has been predicted recently to be occupied by a single electron due to a strong Coulomb repulsion.\cite{Lin} As such, in contrast to shallower donor-electron states with much larger Bohr radius (typically between 5 and 10~\AA) observed in some other band-gap insulating oxides,\cite{Coey} this in-gap state is well localized\cite{Lin} and can not provide extended magnetic coupling in a low-doping limit. Moreover, the effective coupling mediated by the donor electrons is always ferromagnetic, because it is quadratic in the dopant--to--donor-electron magnetic interaction. This contradicts our experimental observations that reveal antiferromagnetic interaction between Mn ions contributing to the broad C3 ESR component.

We are therefore inclined to the scenario of dopant aggregation, where the enhanced ESR line width of the C3 component can be explained by an interplay of the dipolar and superexchange interactions within dense Mn-rich cluster of \STO. With increasing the doping concentration, the fraction of the isolated Mn$^{4+}$ ions on the B sites decreases (see the C2 component in Fig.~\ref{fig5}), as one would expect. Moreover, the decreasing line width with increasing doping suggests that the exchange-narrowing mechanism is becoming more effective. 
Although ESR is not best suited for quantifying the cluster sizes, like some other nano-characterization tools,\cite{Dietl,Bonanni} this experimental finding is a likely fingerprint of nanosized clusters. Namely, for small clusters, an average exchange coupling of a particular Mn$^{4+}$ ion to its neighbors, being a scalar sum, is more affected by nonmagnetic Ti$^{4+}$ neighbors than the average dipolar interaction, resulting from a vector sum.
This effect would, however, die out in bigger clusters where the percentage of the moments on the cluster surface becomes small compared to the moments residing inside the cluster. Thus, the scenario of a bulk phase segregation can safely be dismissed. The situation rather resembles other diluted magnetic semiconductors, where nano-organization of dopants within the host matrix has been detected.\cite{Dietl,Bonanni}

The apparent systematic mismatch between the relative ESR intensity of the C1 component $I_1$ (see Tab.~\ref{table2}) and the site occupancy determined by x-ray absorption near-edge structure (XANES) spectroscopy\cite{Valant} (see Tab.~\ref{table1}) indicates that also Mn$^{2+}$ ions at the A sites aggregate, although only partially.  Since the Mn[Mn]O$_3$ cubic perovskite structure is unstable and a perovskite-type structure stabilizes only at high pressures and high temperatures,\cite{Ovsyannikov} it would be structurally highly unfavorable if both Mn$^{2+}$ and Mn$^{4+}$ ions entered the same clusters. 
Therefore, a formation of separate MnTiO$_3$ and SrMnO$_3$ nanoclusters is expected. We note that the former compound in bulk adopts the ilmenite crystal structure\cite{Stickler} that is significantly different from the cubic perovskite structure of \STO~as the Mn$^{2+}$ ions reside in oxygen octahedra. SrMnO$_3$, on the other hand, can form a cubic perovskite polymorph.\cite{Sondena}
This then probably explains the different tendency of Mn$^{2+}$ and Mn$^{4+}$ to aggregate. 
Quite importantly, we point out that the percentage of the aggregated Mn$^{2+}$ ions in \STO~is correlated to the sample quality. There is a considerably larger mismatch between $I_1$ and the A-site occupancy,\cite{Valant} i.e., higher level of Mn$^{2+}$ aggregation in the A3N$_2$ than in the A3O$_2$ sample, while, on the other hand, the presence of structural defects determined by transmission electron microscopy (TEM) is much lower in A3N$_2$, as its microstucture is more uniform.\cite{Valant}

Last, we return to the $\mu$SR results. Again, the homogeneous picture of the donor-electron-mediated exchange can not explain the rather slow fluctuation rates of local magnetic fields at the muon stopping sites (see Tab.~\ref{table1}). The derived fluctuation rate $\nu = 20$~MHz is orders of magnitude below the expected rate for exchange-coupled moments, $k_B J/h\sim60$~GHz ($J=3$~K). In the inhomogeneous picture of clusters, on the other hand, the volume fraction of such clusters will be very small for the low doping concentrations of our samples. Therefore, the muons will predominantly couple to the diluted isolated moments outside the clusters, which fluctuate due to much smaller dipolar interactions.
A small fraction of muons that do stop inside or close to clusters is likely responsible for the overestimation of the predicted LF decoupling at intermediate fields [Fig.~\ref{fig1}(c)].
 
\section{Conclusions}
In conclusion, our muon spin relaxation and electron spin resonance investigations of Mn-doped \STO~have demonstrated unambiguously that this system, independent of the nominal doping site and post-treatment atmosphere, intrinsically remains paramagnetic down to low temperatures. Both, $\mu$SR depolarization curves as well as ESR spectra failed to detect any irregularities in the A-site-doped samples at the presumed spin-glass transition temperature around 40~K. 
In addition to several recent reports suggesting an extrinsic nature of ferromagnetism in many DMOs, our study, disproving the intrinsic spin freezing in the Mn-doped \STO~materials, serves as a warning that the phase-segregation processes in not optimally processed materials can lead to very faint compositional inhomogeneities that are far below the limits of conventional analytical techniques.
Moreover, we have discovered site-dependent aggregation of the dopants in Mn-rich clusters within the \STO~matrix. Quite importantly, such spinodal decomposition is correlated to the quality of the sample's microstructure. Moreover, it is less favorable for the Mn$^{2+}$ ions that partially remain isolated in diluted regions, as revealed by the
remaining ESR sextet. Finally, our study has also unraveled the origin of the dominant broad ESR component commonly observed in the ESR studies of the Mn-doped \STO~powders and crystals. This component emerges from the Mn-rich clusters of \STO~and is due to exchange coupling and dipolar interactions between dopants.

\acknowledgments
This work has been supported by the Slovenian Research Agency Programs P1-0125, P2-0377 and Projects N2-0005, Z1-5443. The $\mu$SR part of this work is based on experiments performed at the Swiss Muon Source (S$\mu$S), Paul Scherrer Institute, Villigen, Switzerland.

\appendix

\end{document}